\documentclass[12pt]{iopart}

\usepackage{breqn}
\usepackage{iopams}  
\usepackage{graphicx}
\providecommand{\keywords}[1]{\textbf{\textit{Index terms---}} #1}
\begin{document}

\title{First Results of GINGERino, a deep underground ring-laser}

\author{J. Belfi$^1$, N. Beverini$^{1,2}$, F. Bosi$^3$, G. Carelli$^{1,2}$, D. Cuccato$^{3,4}$, G. De Luca $^5$, A. Di Virgilio$^1$, A. Gebauer$^{6,7}$, E. Maccioni$^{1,2}$,  A. Ortolan$^3$,
 A. Porzio$^{8,9}$, R. Santagata$^{1,10}$,  A. Simonelli$^{1,7}$,  and  G. Terreni$^1$,}
\address{$^1$  INFN Sezione di Pisa, Pisa, Italy}           
\address{$^2$   University of Pisa, Pisa, Italy}            
\address{$^3$ LNL Laboratorio INFN, Legnaro, Italy}  
\address{$^4$ DEI, University of Padova-Italy}
\address{$^5$ INGV  Centro Nazionale Terremoti, L'Aquila, Italy}
\address{$^6$ Technische Universitaet Muenchen,  Munich, Germany}
\address{$^7$ Ludwig-Maximilians-University Munich, Germany}
\address{$^8$  INFN Sezione di Napoli, Napoli, Italy}
\address{$^9$  CNR-SPIN Napoli, Napoli, Italy}
\address{$^{10}$   University of Siena, Siena, Italy}

\ead{angela.divirgilio@pi.infn.it}
\vspace{10pt}
\date{ }

\begin{abstract}
Large ring--laser gyroscopes are capable of measuring angular rotations with a precision well below fractions of $prad/s$, not far from 
$10^{-14}$ $rad/s$, the accuracy required for General Relativity tests, that is what the GINGER
(Gyroscope-IN-GEneral-Relativity) experiment, based on an array of ring-lasers, is aiming at.
The $\it{ring-laser} $ is now a mature technique: it has 
high sensitivity and very large bandwidth, it allows continuous data taking, the apparatus can be 
actively controlled for obtaining measurement times of the order of months so to obtain reliable low 
frequency informations.
These features, alone, do not guarantee the possibility of measuring the General Relativity Lense--Thirring effect that manifests itself as a tiny ($\approx 10^{-9} \times \Omega_E$) steady state perturbation of the Earth rotation rate.
The main question is whether a global quantity, as the Earth rotation rate can be measured locally by an instrument based on ring--laser with such an unprecedented accuracy.
To this end, care is necessary in order to guarantee that the measured quantity is not affected by local disturbances.
An underground location being in principle less affected by external local disturbances as hydrological changes, barometric pressure and temperature variations, represents a good candidate for housing such a challenge.
GINGERino is a test apparatus to investigate the residual local disturbances in the most inner part of the underground international laboratory of the GranSasso (LNGS). It consists of a square ring laser with a $3.6$ m side. Since larger rings imply higher sensitivity, the instrument has been tailored to be the larger allowed by the particular location inside the laboratory.
Its main objective is to measure the very low frequency rotational motions, in order to prove, or not, that LNGS is a suitable location for very low noise measurements and, possibly, General Relativity tests.
Aside this main goal, GINGERino will provide unique data for geodesy and geophysics. Its installation has been completed during 2015. Since then, some data set of several days of continuous Earth's 
rotation measurements have been collected, with the apparatus running unattended. The typical power spectrum sensitivity was a few $ 10^{-10} rad/s/\sqrt(Hz)$, with integration time not longer than tens of seconds. Improvements of the apparatus are ongoing in order to improve the integration time.
This paper aims at presenting the first results obtained by GINGERino and describing its concept with all the strategies implemented for reaching higher accuracy and long term stability.
\end{abstract}
\pacs{....}
\\
\submitto{\CQG}
\maketitle                           
\section{Introduction}            
Ring laser gyroscopes (RLG) \cite{RSIUlli} are, at present, the most precise sensors of absolute angular velocity for an Earth based apparatus. 
They are based on the Sagnac effect arising from a rigidly rotating ring laser cavity.
They are essential in estimating rotation rates relative to the local inertial frame in many contexts ranging from inertial guidance to angle metrology, 
from geodesy to geophysics. The Gross Ring "\textbf{G}" at the Wettzell Geodetic Observatory has obtained a resolution on the Earth rotation rate of
$3 \times 10^{-9}$ (about $15\times 10^{-14}$ $ rad/s$ with 4 hours integration time) \cite{RSIUlli, EarthUlli}.
Such an unprecedented sensitivity shows that this class of instrument is suited to probe the spatio-temporal structure of the local gravity field.
Earth rotation is, actually, precisely measured by  an international system of very long baseline interferometers (VLBI). VLBI has demonstrated highly accurate 
and stable determinations of the universal time, in large part because its very precise
observations of extragalactic radio sources provide access to
a nearly inertial celestial reference frame. A large RLG is an instrument whose output is directly linked to the instantaneous axis of rotation of the Earth. 
Furthermore one obtains a continuous set of measurements, which is not yet available for VLBI. A good  agreement between VLBI and ring laser has been 
achieved by measuring and comparing the low frequency Chandler-- and Annual-- Wobble, a free oscillation of the Earth \cite{ChandlerUlli}. 
The remaining discrepancy is due to the fact that local tilts of \textbf{G}, a single component ring laser, currently cannot be measured with a sufficient 
long-term stability. An improvement in sensitivity and a full 3-dimensional detection of the Earth rotational velocity vector would allow RLG to integrate efficiently 
the data produced by VLBI and, possibly, measure the Lense--Thirring effect. 
GINGER (Gyroscopes IN GEneral Relativity) will aim at measuring the gravito-magnetic (Lense--Thirring) effect of the rotating Earth by means of an array of high sensitivity and accuracy ring lasers. 
In the weak--field approximation of Einstein's equation, the  response $\nu$ seen by a RLG located in a 
laboratory on the Earth surface, with co--latitude $\theta$,  and with the axis contained in the meridian plane at an angle $\psi$  with respect to the zenith, can be written as: 
\begin{center}
\begin{equation}
\fl \nu = 4\frac{S}{\lambda} \Omega_E[\cos{(\theta +\psi)}-2\frac{GM_E}{c^2R_E}\sin{\theta}\sin{\psi}+
   \frac{GI_E}{c^2R_E^3}({2\cos{\theta}\cos{\psi}+\sin{\theta}\sin{\psi}})]
 \end{equation}
\end{center}
where $S$ is a geometric scale factor for the RLG, $G$ the Newton's gravitational constant, $\Omega_E= 7.29\times10^{-5}$ $rad/s$ the Earth's instantaneous angular rotation speed, $M_E$ the Earth's mass, $R_E$ the Earth's mean radius, and $I_E \approx \frac{2}{5} M_E R_E^2$ the Earth's moment of inertia. The first term corresponds to the standard Sagnac signal; the second one, known as geodetic or De Sitter precession, is produced by the motion of the laboratory in the
curved space--time around the Earth and, the third one, known as Lense-Thirring precession (LT) and characterised by a dipolar structure, is produced by the rotating mass of the Earth and is proportional to the Earth angular momentum \cite{NoiPRD}.
The last two terms are both relativistic, but their contributions can be discriminated by a vectorial reconstruction of the rotation speed. They are smaller than the classical Sagnac effect by a factor of $\approx 10^{-9}$, that is of the order of magnitude of the ratio between $R_E$ and the Schwarzschild radius of the Earth  $2GM_E/c^2$.
These effects should be observed as a difference between the rotation rate observed by the array of RLGs in the rotating frame of the laboratory, and the length of the day determined in the "fixed stars" inertial frame by IERS (International Earth Rotation and Reference System) through VLBI. 
Registering a perturbation that amounts to 1 part in a billion of the Earth rotational rate, requires an unprecedented sensitivity of the apparatus. 
An array of at least three ring lasers would allow us to vectorially measure the Earth's angular velocity and, having at disposal the time series of the daily estimate of Earth rotation vector from the IERS Service 
(http://www.iers.org), it would be possible to isolate the Geodetic and Lense--Thirring contributions.  Moreover, 
as it has even been discussed in recent papers \cite{NoiseUlli},
 the external disturbances are a severe  limitation, if the apparatus is located on the Earth surface,.
An underground location, far from external disturbances as hydrology, temperature and atmospheric pressure changes, 
is essential for this challenging experiment, and LNGS (Laboratori Nazionali del GranSasso,  the underground INFN National laboratory) may be a suitable location.
 
%

LNGS underground laboratory exhibits a very high natural thermal stability and, being deep 
underground, it is not affected by top soil disturbances. Moreover, being a very large
laboratory, it seems feasible, if needed, to further shield GINGER in order to reduce the anthropic disturbances. 
In order to check how much the underground location is an advantage for GINGER, a single axis apparatus called GINGERino, has been installed inside LNGS. This installation is a prototype dedicated 
to GINGER and to the utilisation of RLG for fundamental science, but at the same time provides unique information for geophysics \cite{simonelli}.
When the sensitivity will be good enough it should also provide local measurements relevant for geodesy and geophysics, especially in parallel observations with other large RLG like \textbf{G},
as daily and semidiurnal polar motion.  The construction of GINGERino has been completed by the end of 2014, and it has been taking data for the first time during the spring, and in October 2015. In the following the apparatus will be described, and  the very preliminary data will be reported and discussed. In the conclusions the near future development will be sketched.
\section{The GINGERino Apparatus}               
The whole installation consists of a 3.6 m in side square ring-laser and few high sensitivity co-located geophysical instruments.
These are tilt meters with nrad resolution (2-K High Resolution Tiltmeter (HRTM), Lipmann) and high performance seismometers (Trillium 240s and Guralp CMG 3TÐ360s). This combination of different instruments will improve the knowledge of the
behaviour of the location, and will be essential in the interpretation of geophysical data. In the following
the different components will be described together with data acquisition and analysis.
\subsection{The ring-laser and the granite monument}
\begin{figure}
 		\centering
		\includegraphics[scale=0.08]{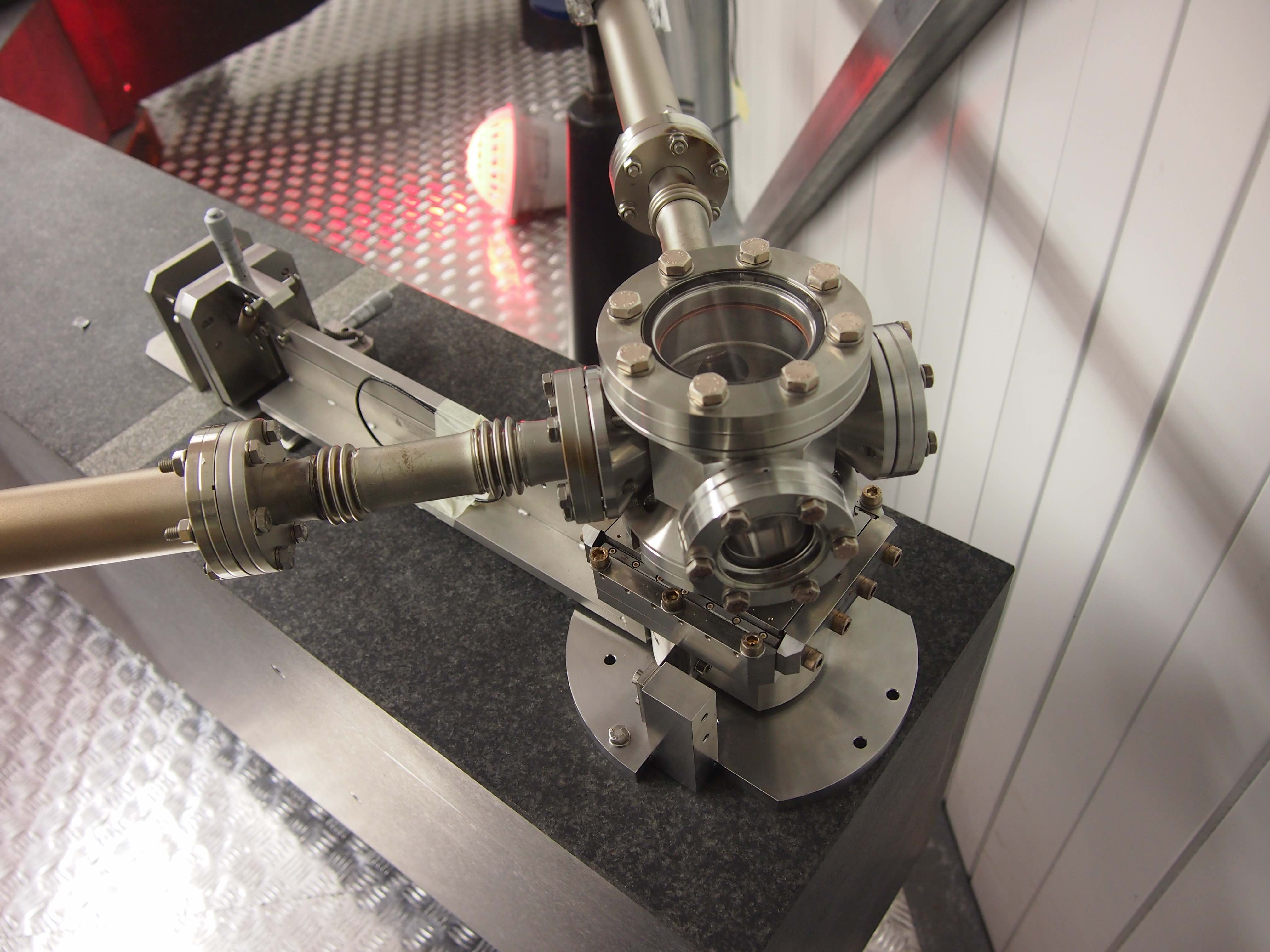} 
 		 		\caption{Detail of one of the four mirror boxes, the two output viewports and the micrometric system to tilt the mirrors are well visible.}
 		\label{fig1}
 	\end{figure}

The Gross ring \textbf{G} is based on a monolithic mechanical design which cannot be extended to form an array. In order to circumvent this engineering limitation, hetero-lithic structure designs 
have been developed in the last decades. The first prototype of this kind was a GeoSensor, which was developed for seismology, were the long term stability of the apparatus is not an issue. In order to obtain with the hetero-lithic RLG the same long term stability of the monolithic G it is necessary to develop a suitable active control scheme. For that purpose we have developed a new prototype, called GP2 (actually running at INFN Pisa) \cite{NoiCR, PhDRosa, NoiRosa}, which is a test bench to test the geometry control scheme. 
Our first prototype G-Pisa (no more in operation) \cite{NoiAPB, NoiVirgo} was based on a modified GeoSensor design.
In this first installation, GINGERino (see Fig. \ref{fig1}.) uses the mechanical parts of G-Pisa; in the future, the GP2 prototype could be replaced. It is made up of $4$ mirror boxes connected by vacuum pipes. 
Each mirror can be independently moved, with sub-micrometer resolution, so to align the optical cavity. 
Two piezoelectric translators can be used to stabilise the perimeter in order to compensate for the thermal expansion of the cavity, 
avoiding laser mode hopping and increasing the device duty cycle. The perimeter active control of the prototype G-Pisa has been successfully tested in the past \cite{NoiVirgo}.

 
 \begin{figure}
 		\centering
		\includegraphics[height=8cm,width=7cm]{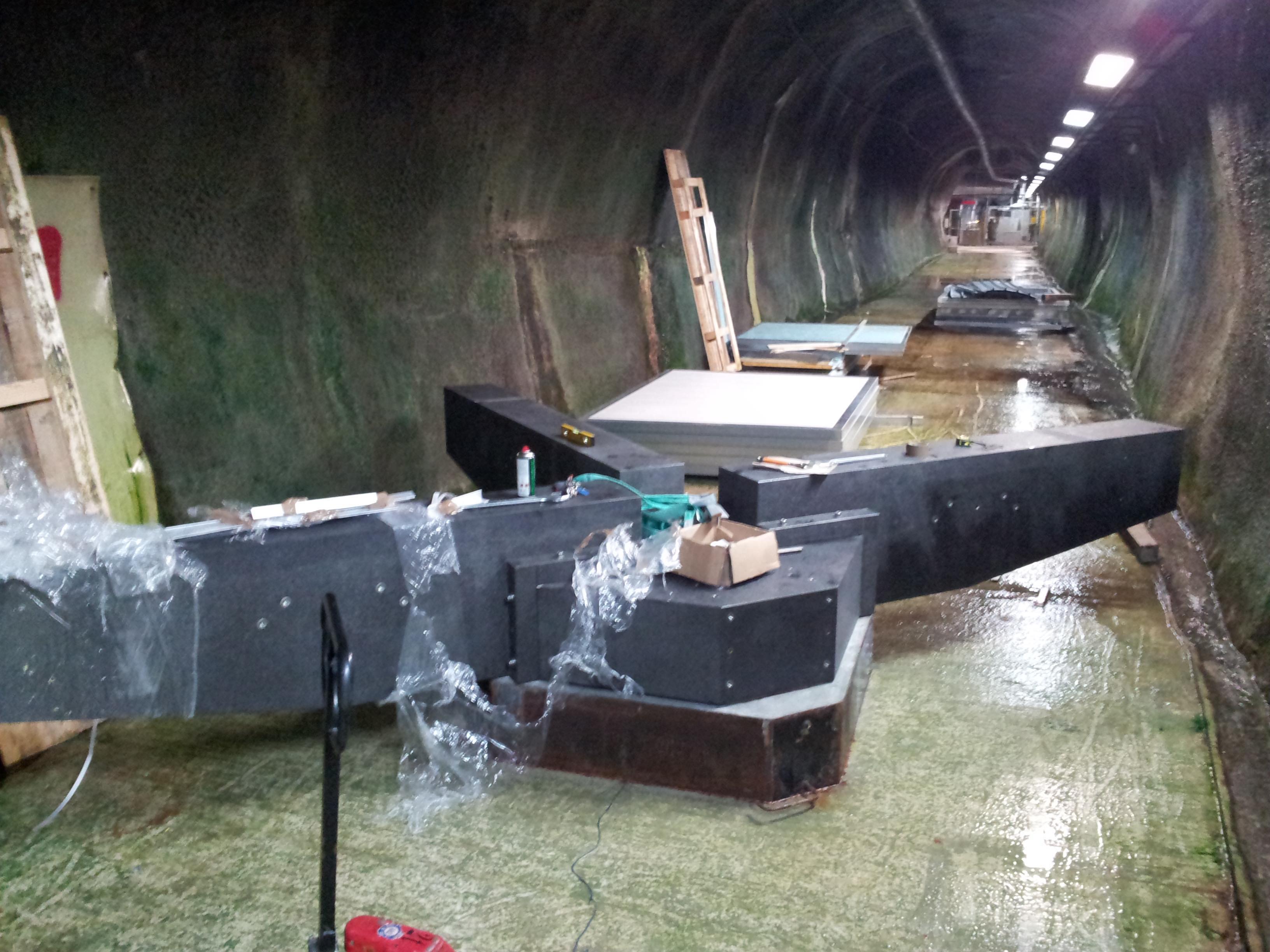} 
 		 \includegraphics[scale=0.2]{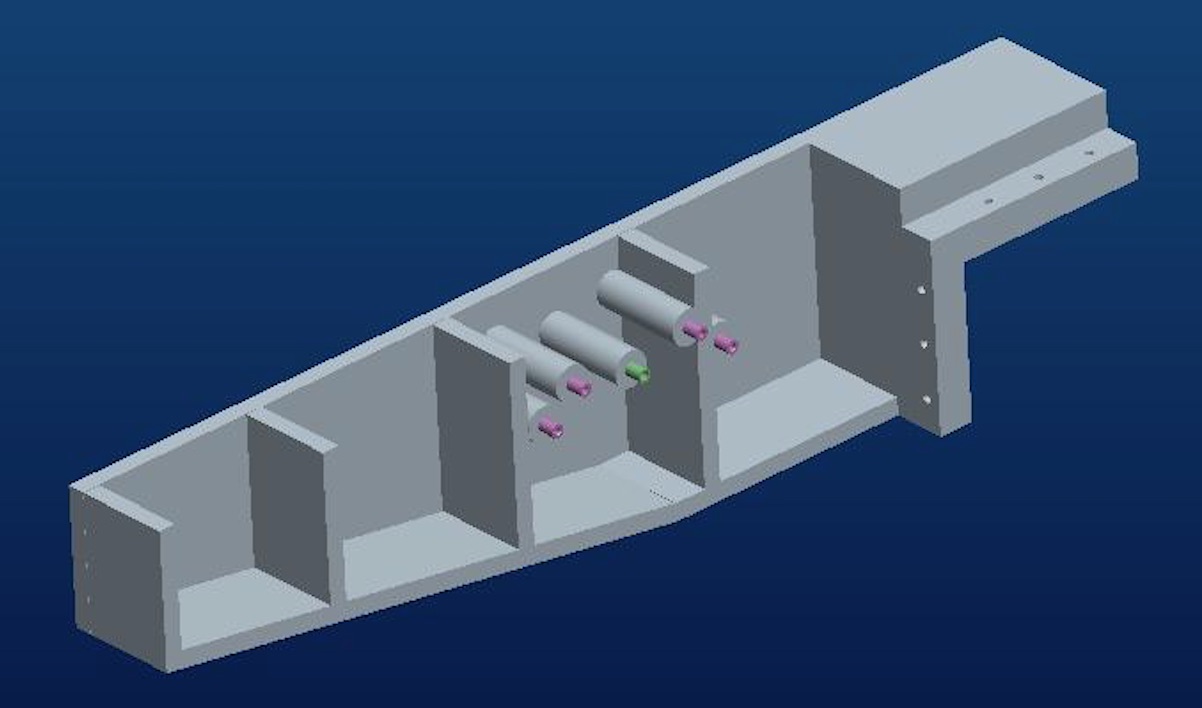}
 			\caption{Left: September 2014, the central block is attached to the concrete interface, and three arms are in place.
 					Right: a drawing which shows the inner part of the arms, made by lightened granite, i.e. slices of granite 
 					glued together and machined.}
 				\label{construction}
 		\end{figure}	
 

%

The ring-laser is tightly attached to a cross structure made of black African granite, composed by a central octagonal massive block (3 tons), 
and four lightened arms each weighting $\approx 800\, kg$ (see Fig. \ref{construction}). The granite structure is screwed to a reinforced concrete block integral to the underneath bedrock. The African black granite has been chosen because it can be machined with high precision and has quite a low thermal expansion coefficient ($7\times10^{-6}$ $/^oC$).

 

\begin{figure}
 		\centering
 		\includegraphics[scale=0.1]{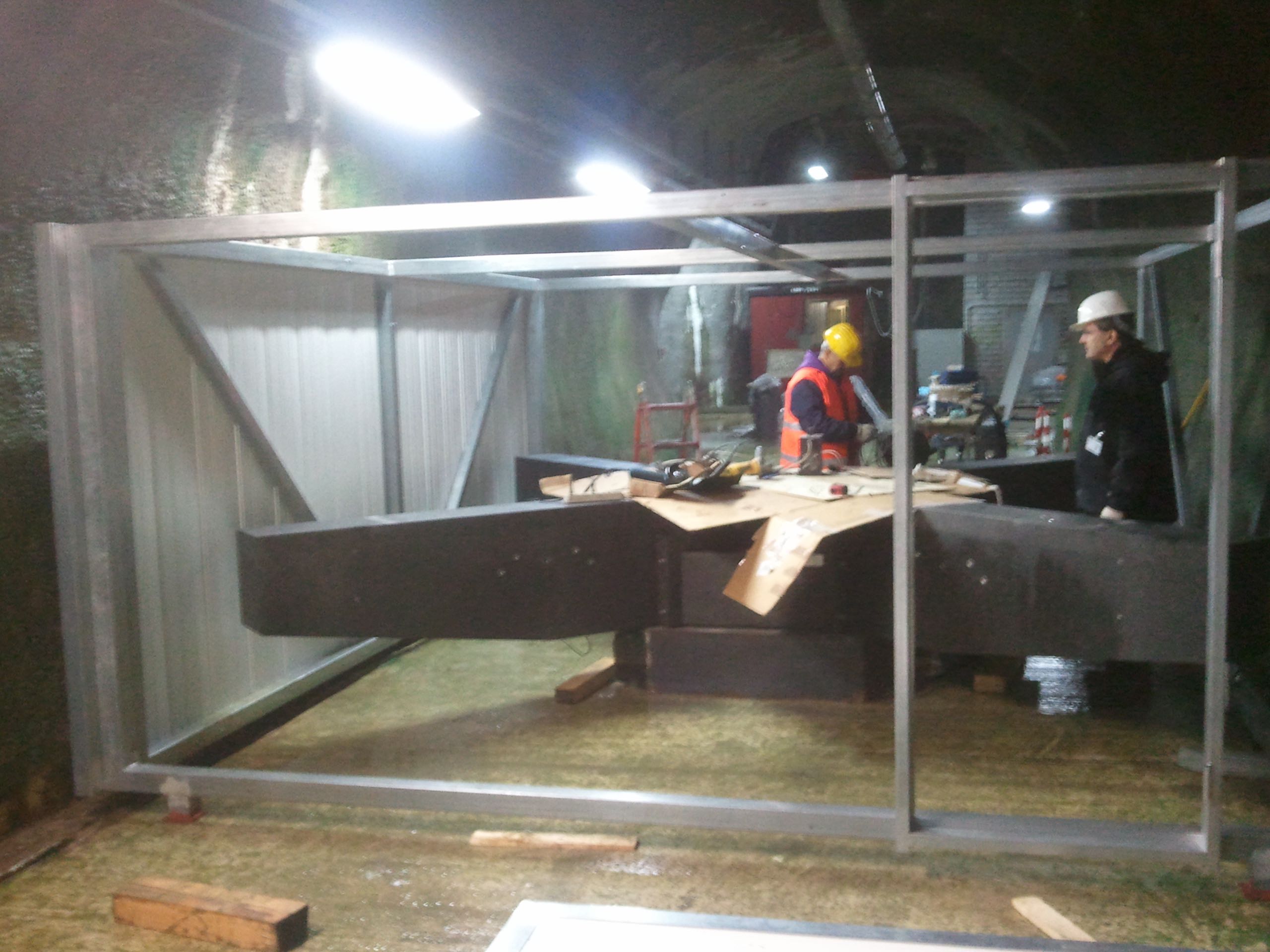} 
		\includegraphics[scale=0.08]{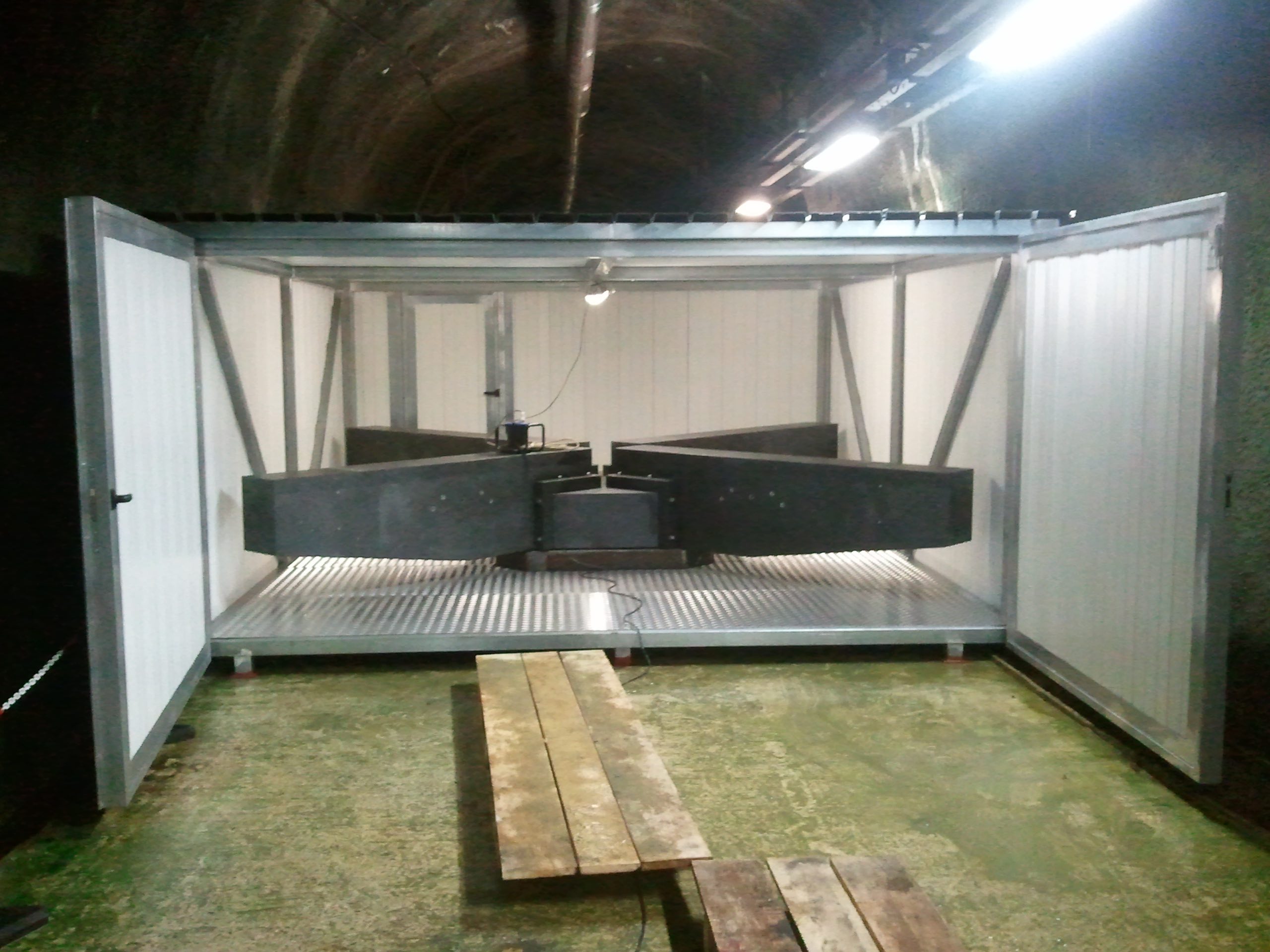}   
 		\caption{Top:  Early phase of the installation of the chamber,  Bottom: the chamber is complete}
 		\label{box}
 	\end{figure}

 The advantages of a single support for the mirrors are: a) a better definition of the geometry and planarity since the granite can be very precisely machined; b) the whole set-up is attached to its center and the whole granite cross is inside the same thermal bath.
The installation area was at a temperature of $7$ $^oC$ with a relative humidity close to the dew point all the year round. The whole installation is now protected by a large anechoic box (see Fig. \ref{box}). Infrared lamps are used to increase the temperature inside the box thus reducing the relative humidity from more of $90\%$ down to $\approx 50-60\%$. 
So far, this infrastructure has been running for several months, and has shown that it keeps the GINGERino area at a 
temperature around $14-18$ $^oC$. We will investigate later if this system needs to be improved with additional shielding and/or an active temperature control, in order to improve the long term stability of the temperature. 
 	
\subsection{DAQ system, Online and Control}
Our instrumentation runs unattended, in this way vibrations and temperature changes induced by the human presence are minimized.
Through the DAQ system, not only the data to be analysed off-line are acquired, but as well the status of the apparatus can be monitored, the mirrors can be moved by means of piezo actuators, and, if necessary, the control loops can be opened.
Sometimes, it is necessary to remotely move one of the mirrors to change the operating condition of the laser. 
The DAQ system itself is remote-controlled and transfers the data from INFN-LNGS to INFN-Pisa. The hardware has been selected in order to be transportable; its main features can be listed as follows:
\begin{itemize}
\item analog to digital conversion and storage of the Sagnac and the two mono beams signals from the ring laser, with $5$ kHz sampling rate;
\item analog to digital conversion and storage of environmental signals (temperature, humidity, pressure), laser parameters (plasma intensity, average intensities, piezoelectric transducers driving voltage)
and local tiltmeters (nano-rad precision), with $1$ Hz sampling rate;
\item analog to digital conversion and storage of seismic channels (seismometers/accelerometers) at 100 Hz sample rate;
\item real-time processing of experimental parameters connected to laser gain, backscattering phase, actuators signals required by active control loops;
\item digital to analog generation of the signals driving the laser, necessary for some of the controls of the apparatus;
\item the sampling frequency of the DAQ system is synchronised with GPS, that is necessary for constant sampling the Sagnac frequency. A time stamp, 
with the accuracy of few milliseconds, is used to record seismic events and for comparing the ring laser data with the data of other instruments.
\end{itemize}
DAQ hardware is based on a modular PXI system. A PXI system is composed of three parts: the chassis, the controller and one or more modules.
The acquisition system of GINGERino is based on components by National Instruments.\\
\subsection{Acquisition Software} 
The operative system running on the PXI-8106 controller is LabView-RT, a real-time system provided by National Instruments. 
The development environment chosen for the implementation of the DAQ is LabVIEW graphical programming language by National Instruments.
Software development occurs on a Windows running PC and subsequently transferred to the PXI controller via ethernet and finally executed under LabView real-time.
Acquired data are written in the PXI local hard-disk and stored on an hourly basis. Each hour an acquisition file  of about 300 Mb is created.\\
\subsection{Acquisition Timing}
Both frequency and time accuracy are important since the former affects the estimation of the Sagnac frequency and the latter introduces errors in the time stamping of seismic events.
We receive a GPS-synchronized PPS (pulse per second) signal and we are connected to a local NTP server in order to obtain a time stamp with the required precision.
The frequency accuracy is obtained by disciplining the clock PXI-6653 board to the PPS via the PXI-6682. The error on the time stamp is on the other hand limited by the uncertainty on the NTP, 
which is of the order of a few milliseconds. The time vector $t$ of the acquired data is then given by $t=t_0+n*dt$ where $t_0$ is the time-stamp from NTP, 
$n$ is the sample number, $dt=1/f_{sampling}$ is the time sampling time interval.\\

\subsection{Data transfer and storage}
\begin{figure}
 	\begin{center}
 		\includegraphics[width=7cm]{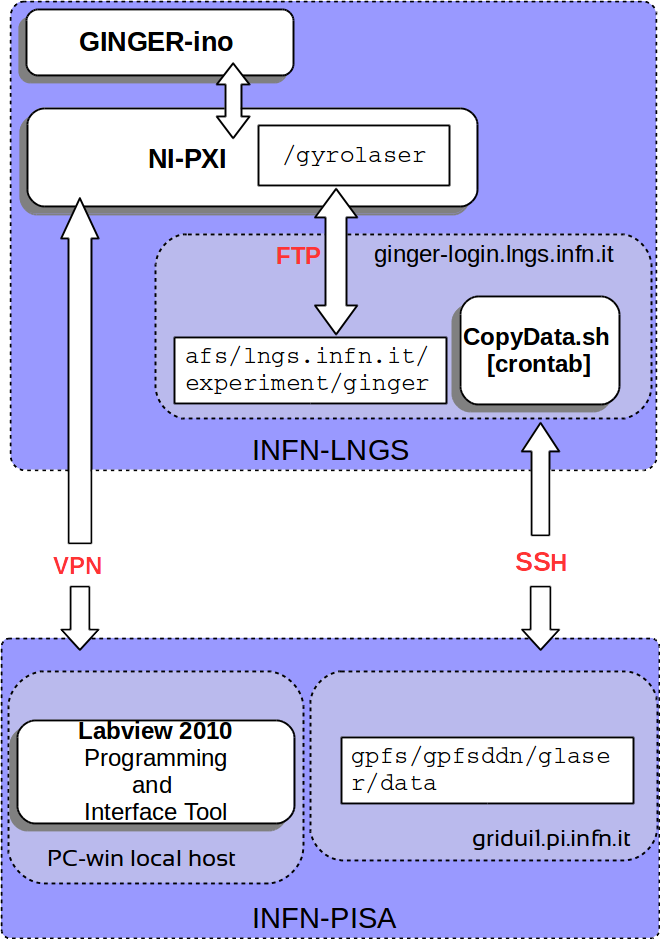} 
 		\caption{Schematics of the connections for the data transfer between INFN-LNGS and INFN-Pisa: VPN and SSH protocols are used for security reasons. }
 		\label{connectionsGINGERino}
 	\end{center}%
 	\end{figure}

The data acquired by the PXI are written on its local hard-disk in a directory containing 1 day of data which is updated hourly in FIFO mode. In this way a buffer of the last 23 hours 
of acquired data is present on the hard-disk for data recovery purposes.
The scheme of the Internet connections between the apparatus at LNGS and Pisa is shown in Fig. \ref{connectionsGINGERino}. The data written on the PXI hard-disk are copied via 
FTP into a dedicated directory on a local virtual machine, that can be accessed from the internet via authorized SSH-account.
The file content of the directory is then copied into the final data transfer destination (at INFN Pisa). 
The data copy service is a cronjob script running every hour on the local virtual machine.
It makes the following operations:
\begin{itemize} 
 \item  compares the file content of the data storage directory on the PXI with the content of the data transfer destination; 
\item  copies  via FTP the missing data files from the PXI to the local storage disk (at INFN LNGS);
\item transfers via SCP from the local storage disk to the data transfer destination.
\end{itemize}
At the end of the process, the data from the PXI are transferred to Pisa, and an image of the PXI buffer is updated hourly on local storage disk. 
\section{Sensitivity of the apparatus}
\begin{figure}
 \centering
 \includegraphics[width=12cm]{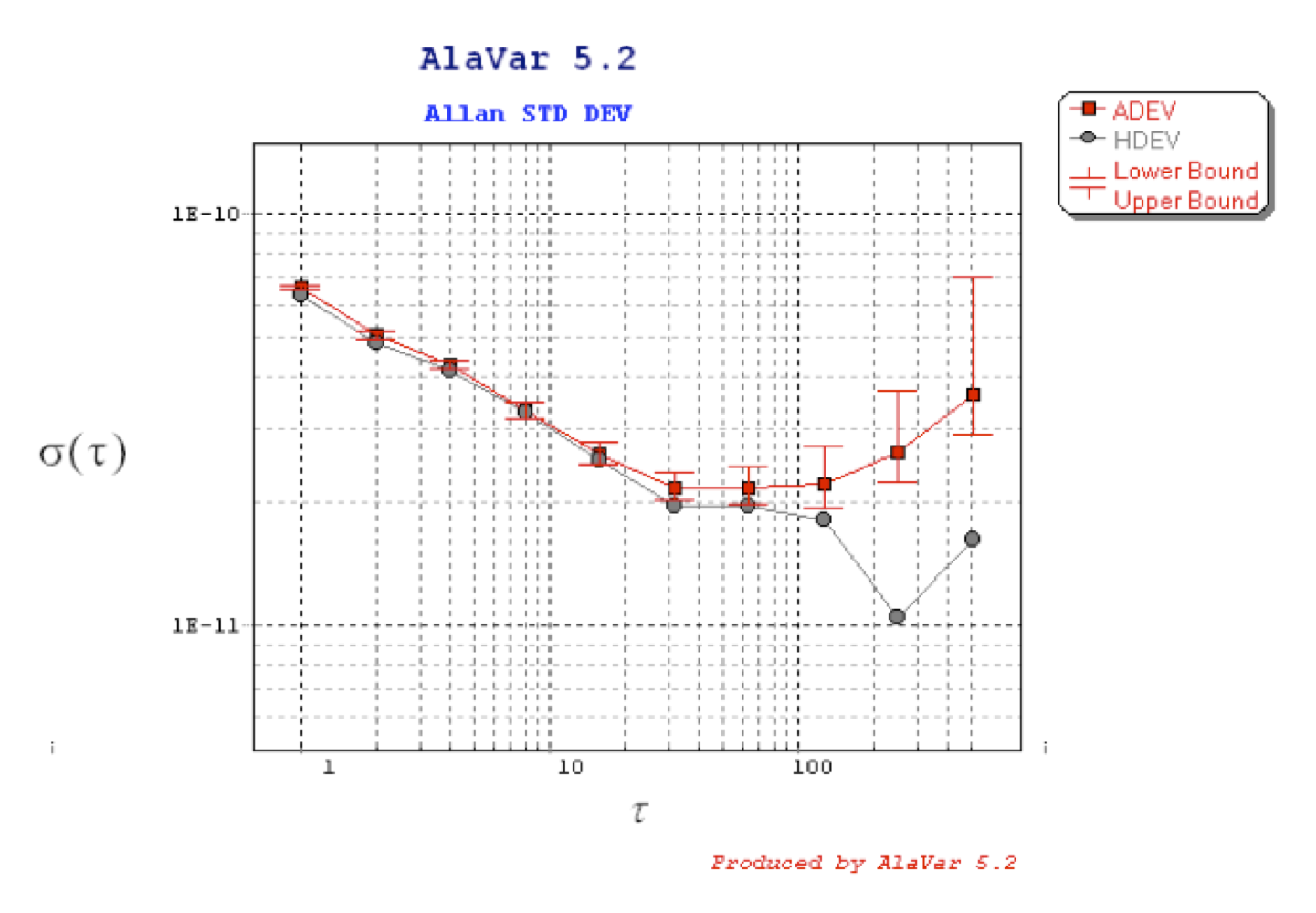} 
 \caption{Allan standard deviation of GINGERino as a function of integration time (no backscattering subtraction), data with reduced backscattering have been selected.  In this picture the Allan is in rad/s, it shows the best sensitivity of $20prad/s$, with $30s$ of integration time.}
 \label{Allan}
 \end{figure}

\begin{figure}
 \centering
 \includegraphics[width=12cm]{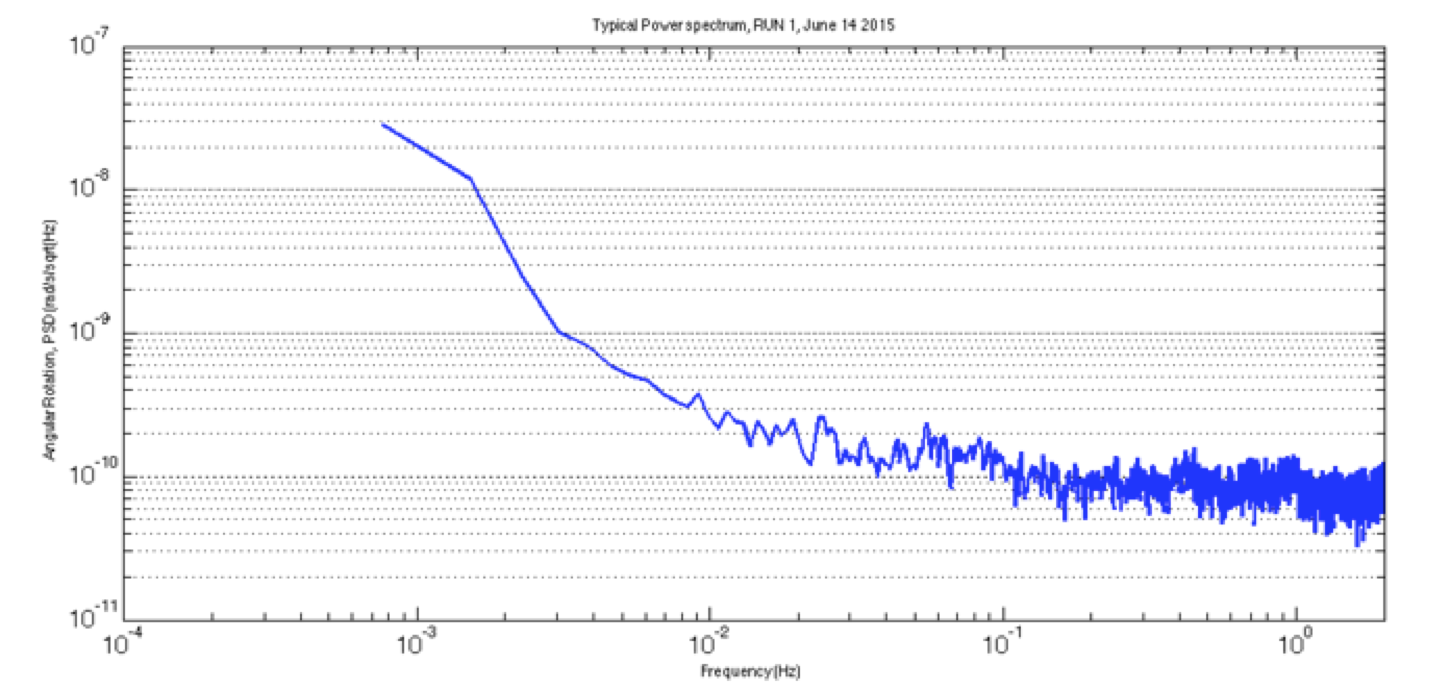} 
 \caption{Angular velocity resolution of GINGERino, directly estimated from the interferogram, as a function of the frequency. The instrument is able to resolve $0.1 nrad/sec/\sqrt{Hz}$ in the range ($10^{-2} - 1$)$ Hz$. In this measurement the information has been filtered of $+/- 2Hz$ around the Sagnac frequency.}
 \label{PSD0}
 \end{figure}

A first set of data have been taken late spring 2015. The measured cavity ringdown time was $\approxeq 250ms$. 
From a direct estimate of the Sagnac frequency by means of the Hilbert transform of the interferogram, 
we deduced an instrumental resolution of $0.1 nrad/sec/\sqrt{Hz} $ in the range $(10^{-2}- 1)$ $Hz$. (see Fig. \ref{Allan}).
As clearly visible in Fig \ref{PSD0}, the long term stability of the instrument was limited to 100 sec, mainly by radiation backscattering on the mirrors.
\subsection{Backscattering analysis}
\begin{figure}
 \centering
 \includegraphics[width=12 cm]{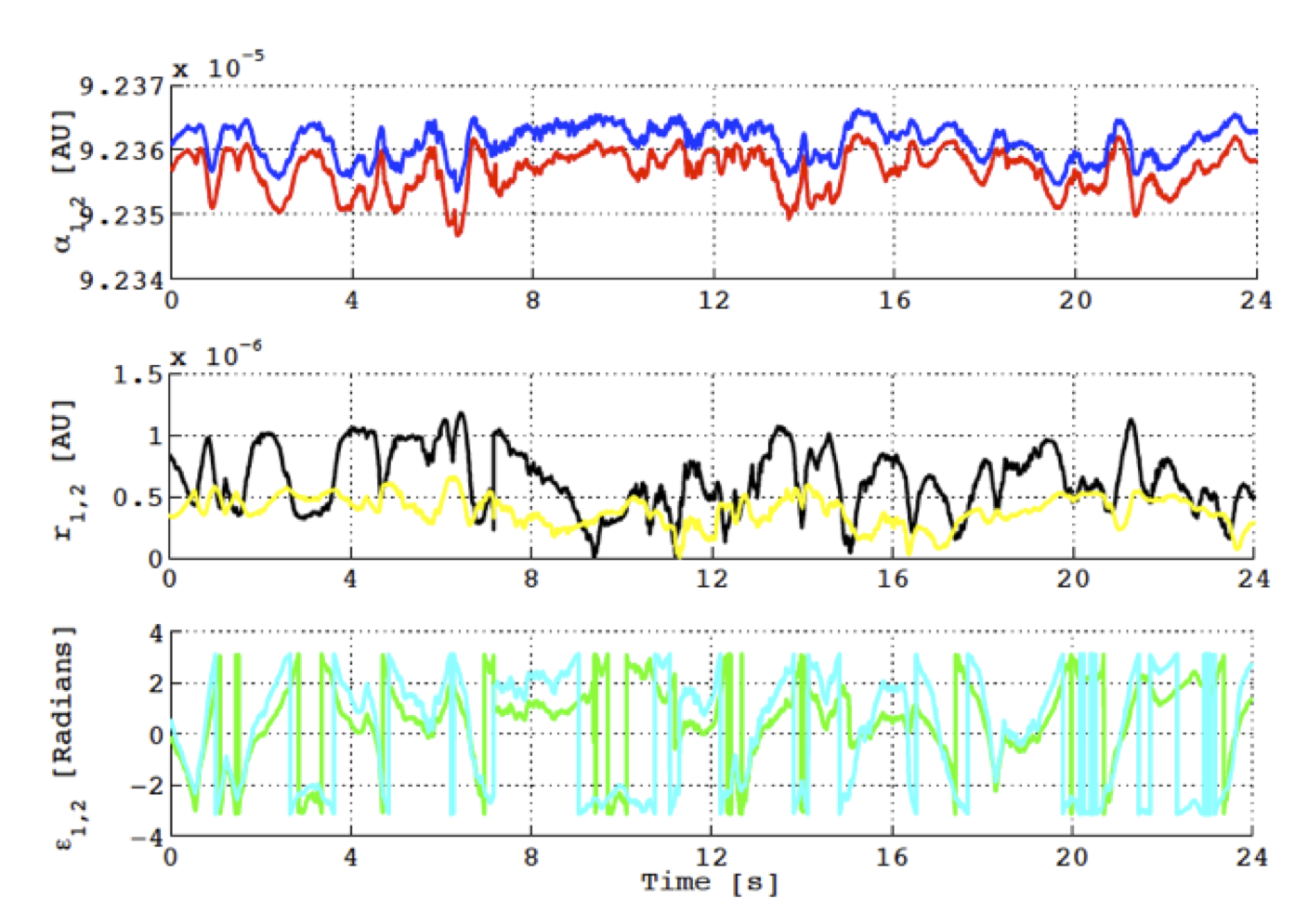} 
 \caption{Lamb parameters describing the laser dynamics for single propagation direction: excess gain minus losses (1 - blue and 2 - red), backscattering amplitudes (1 - black and 2 - orange), and back-scattering phases (1 - cyan and 2 - green). The label 1 and 2 refers to CW and CCW mono-beams, respectively.}
 \label{LAMB}
 \end{figure}

\begin{figure}
 \centering
 \includegraphics[width=12 cm]{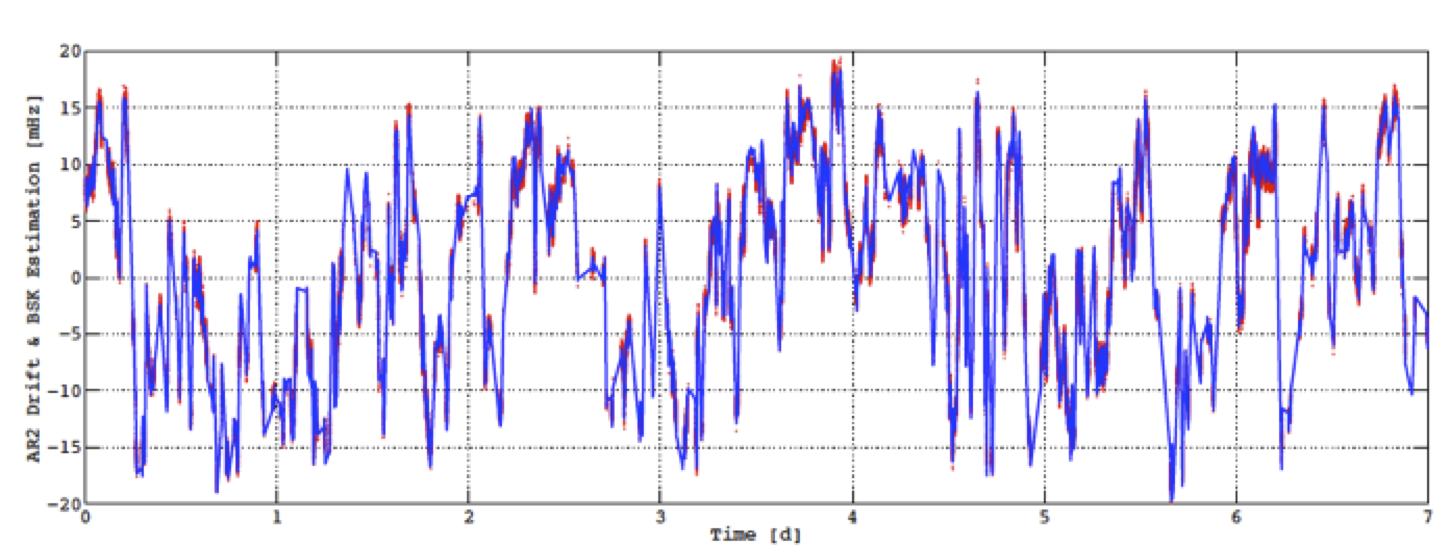} \\
 \includegraphics[width=12 cm]{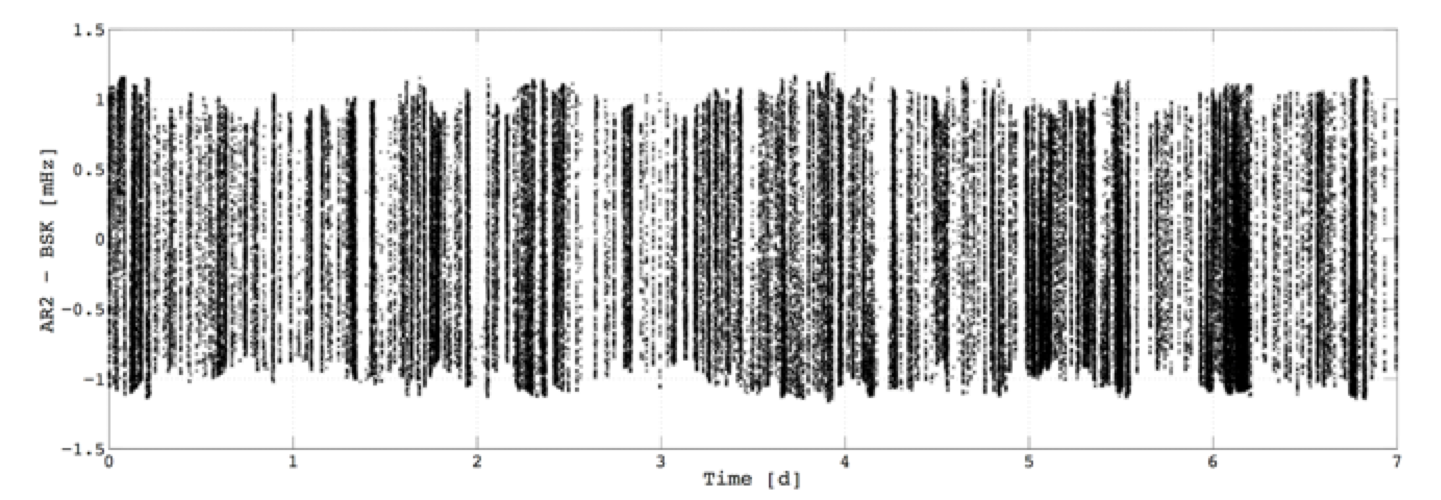} 
 \caption{Top: Comparison of the Sagnac frequency estimated from monobeams  by Kalman filter (blue line), and estimated from the interferogram by AR2 algorithm (red line)
 Bottom: Residuals after the subtraction of the two estimates.}
 \label{topbottom}
 \end{figure}

The strategy for subtracting backscattering noise from ring-laser data  has been extensively discussed in previous papers \cite{NoiMetrologia}, 
where we have shown how and why backscattering noise can be efficiently subtracted, by post-processing the data, applying  Kalman filtering.
The time dependence of backscattering contribution can be also estimated using  a  model which assumes reciprocal ring laser parameters. This approach has been exploited in \cite{Hurst:14}, where the backscattering parameters are estimated by  fitting  amplitudes and phases of the two moonbeam intensities. It has been tested that the two method gives similar results.  
For this analysis purpose several service signals are necessary: the two mono--beams intensities and the laser gain. In this first run the gain monitor was not implemented, 
so we applied backscattering subtraction routine to one week of data, using for the discharge the standard calibration used in 
the past for G-Pisa, and compensating the lack of the gain monitor by adding a free parameter, for details about this procedure, please see [\cite{NoiMetrologia},\cite{CuccatoT}].
The chosen week of data, during the first run of GINGERino, has been from 15th to 21st June 2015 when ring laser run unattended with a 
good sensitivity in the $0.1$ $nrad/sec/sqrt(Hz)$ range.
Data were processed following the procedure already developed for G-Pisa tuning the pre-filters  to the GINGERino Sagnac frequency, 
and estimating the laser parameters by averaging over 10 seconds the mono-beams intensities. Since the single pass gain G was not known, we used the reference value of $G =1.2\times 10^{-4}$ of G-Pisa.
The plots of the identified Lamb parameters are in arbitrary units, except for the back-scattering phase, which is given in radians. 
We firstly extract from the data mono-beam intensities, modulations and phase differences, then we use these quantities to estimate the 
Lamb parameters at a rate of 1 sample every 10 seconds. To give an example, in Fig. \ref{LAMB}, for each counter propagating beam we show the identified laser parameters $\alpha$, excess gain minus losses, $r$, backscattering amplitudes, and  $\epsilon$, backscattering phases,  vs. time in one day of data (16 June 2015).
A preliminary analysis of the data showed that the monobeam intensities and Sagnac frequency are affected by random fluctuations, which are probably due to laser gain variations and cavity deformations.
The selection criterion was based on the correlation between the Sagnac frequency as obtained by AR2 estimate and the same quantity obtained after backscattering subtraction by Kalman filter \cite{NoiMetrologia}.
We set a correlation threshold at $95\%$, thus keeping $13\% $of samples as reported in Fig \ref{topbottom}
 \begin{figure}
 \centering
 \includegraphics[width=12 cm]{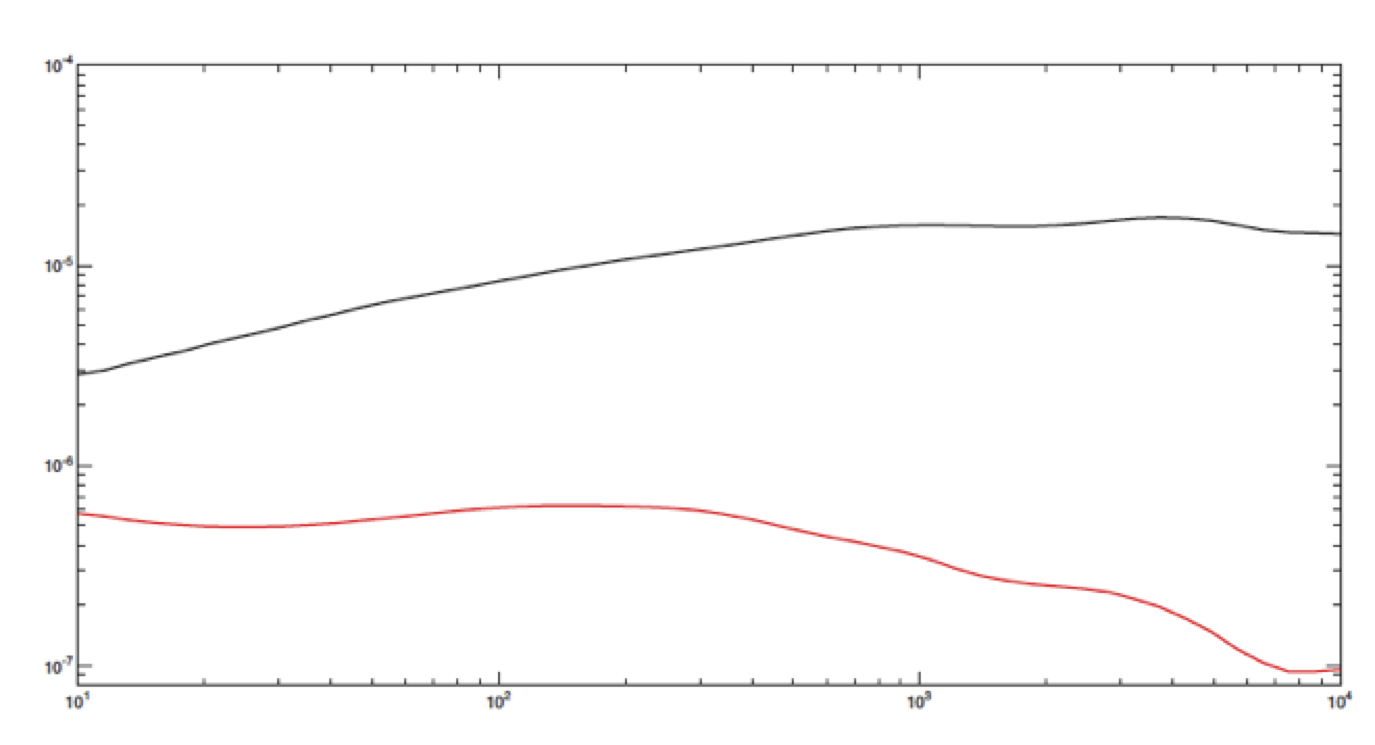} 
 \caption{Allan deviation of the of raw data (black curve) and of the backscattering subtracted data (red curve)}
 \label{LongAllan}
 \end{figure}
The backscattering contribution has been subtracted from the raw data using the correlation coefficient. In this way we overcome the problem of estimating the single pass gain. In Fig.\ref{LongAllan} we show in black  the Allan deviation of the Sagnac frequency before the backscattering subtraction and in red  the same quantity after applying the subtraction procedure.
Notice that the long term stability on time scales greater than 10 seconds has been improved.  Even if Fig. \ref{LongAllan} is preliminary it clearly shows that in one week the average value of the measured Earth rotation rate is stable within $1$ part in $10^7$. It cannot be used to claim that the sensitivity of GINGERino is so high in the long term, but it rather gives the indication that the long term stability of the apparatus is very high, in favour of the installation of GINGER inside LNGS.
\subsection{Few worlds about the mirrors}
The mirrors play a role in the sensitivity, but as well in the backscattering noise, which basically depends on the losses. Ring laser requires top quality mirrors: reflectivity higher than $99.9995\%$, losses as low as possible, possibly  about $1$ $ppm$, transmission not much below $1$ $ppm$. Transmission is an issue, since very little power is leaking out of the ring cavity ( $nWs$), and quite often the mirrors makers in order to reduce the losses reduces as well the transmission. To realise this kind of mirrors top quality substrates, with roughness of the order of fractions of Angstrong, are necessary. This kind of mirrors are not standard, but feasible with dielectric deposition  of thin films realized by very accurate under vacuum deposition. We have as well to take into account that it is very easy to damage this kind of mirrors. In fact, the mirrors of the first run have been damaged during a failure in the vacuum system, a new set has been ordered, but it will be ready not before spring 2016. It is possible to test each mirror before the installation, several suitable techniques have been developed \cite{Isogai13,Ging2013,Sridhar2011} to test high reflectivity and low losses mirrors. Based on this techniques we are developing a test area to characterise each single mirror before the installation inside GINGERino. This test requires a clean room, which is available in the INFN Pisa Section.

\subsection{Analysis of the second run and the seismometers}
\begin{figure}
 \centering
 \includegraphics[width=12 cm]{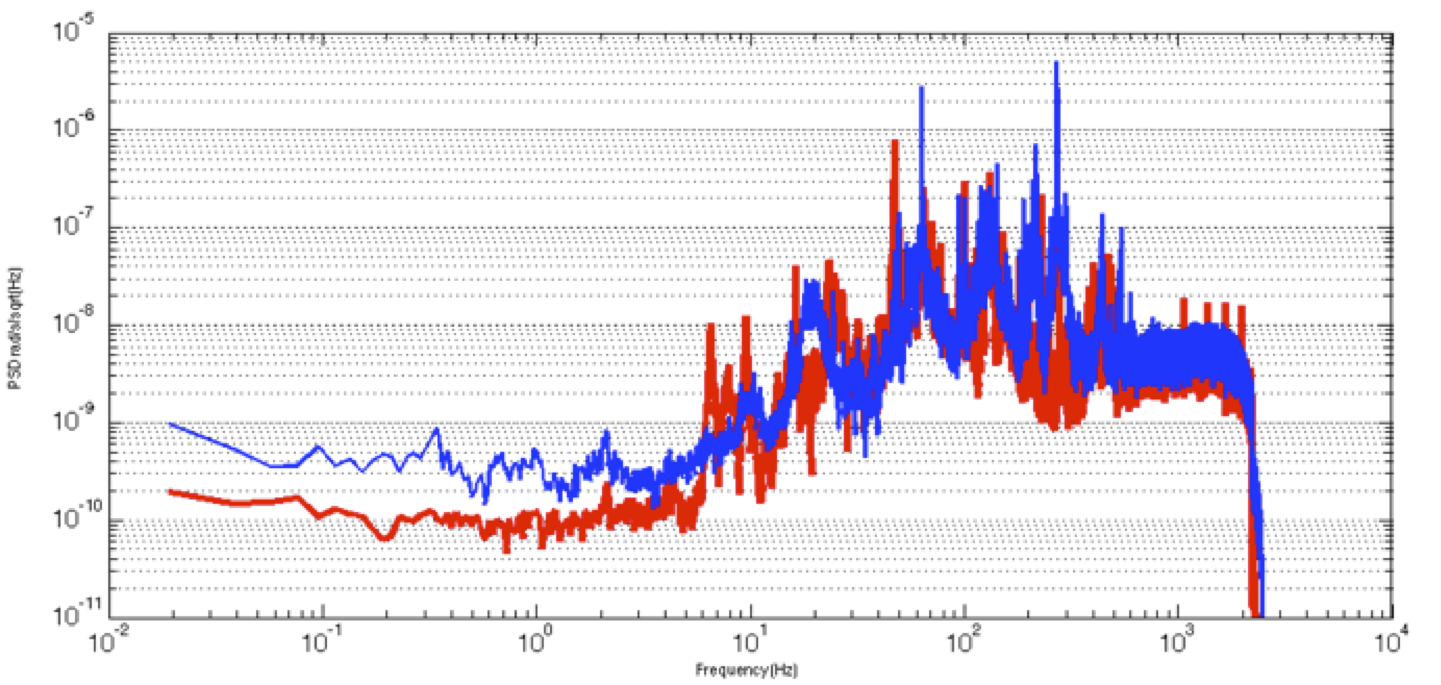} 
 \caption{Typical Power Spectral Densities from runs 1 and 2. The sensitivity, see low frequency, of the new run is not as good as in the first run, this is due to the fact that the mirrors quality is lower. There is an improvement in the high frequency part of the spectrum, due to the improvement on the isolation of the instrument from the floor. (raw data are used, no filters applied)}
 \label{comparison}
 \end{figure}

\begin{figure}
 \centering
 \includegraphics[width=12 cm]{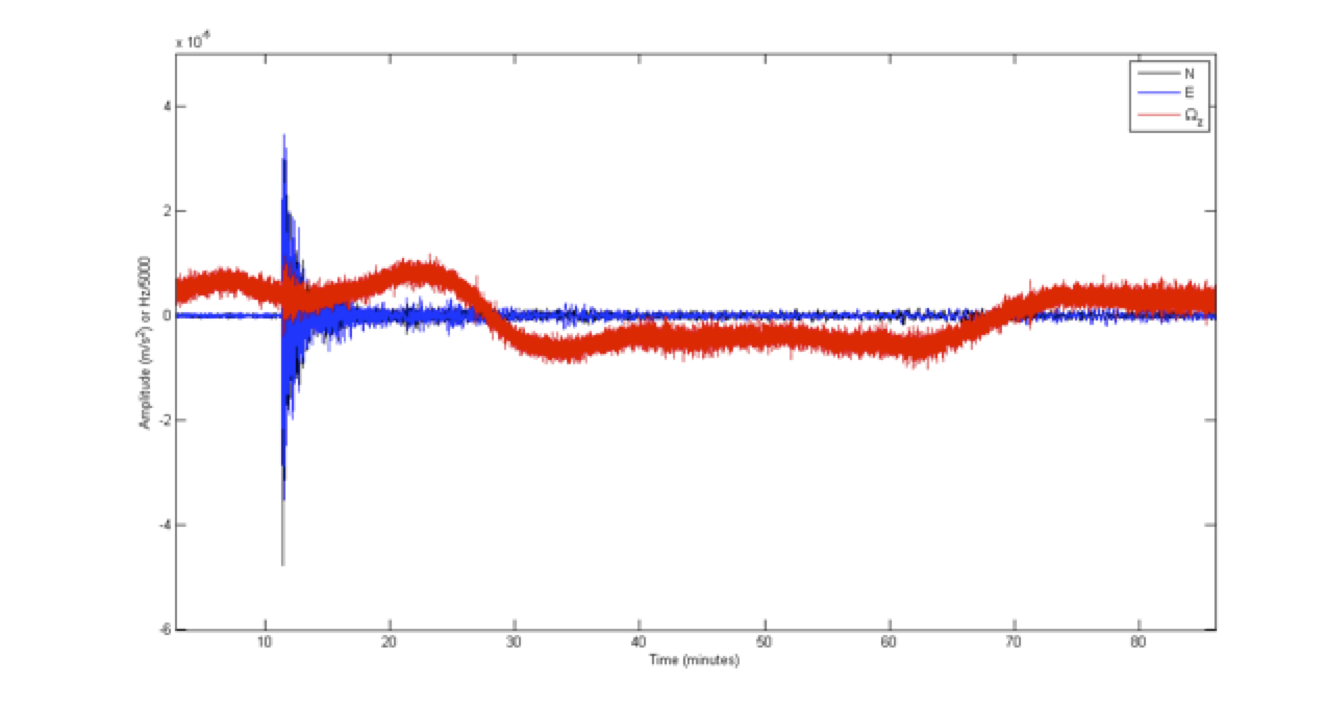} 
 \caption{Teleseismic event of the Vanuatu earthquake. The timing between ring-laser (red curve) and seismometers shows a good synchronization. The two instruments are acquired by two different acquisition systems, in particular the seismometers are acquired by the INGV network GAIA.}
 \label{vanuatu}
 \end{figure}

\begin{figure}
 \centering
 \includegraphics[width=12 cm]{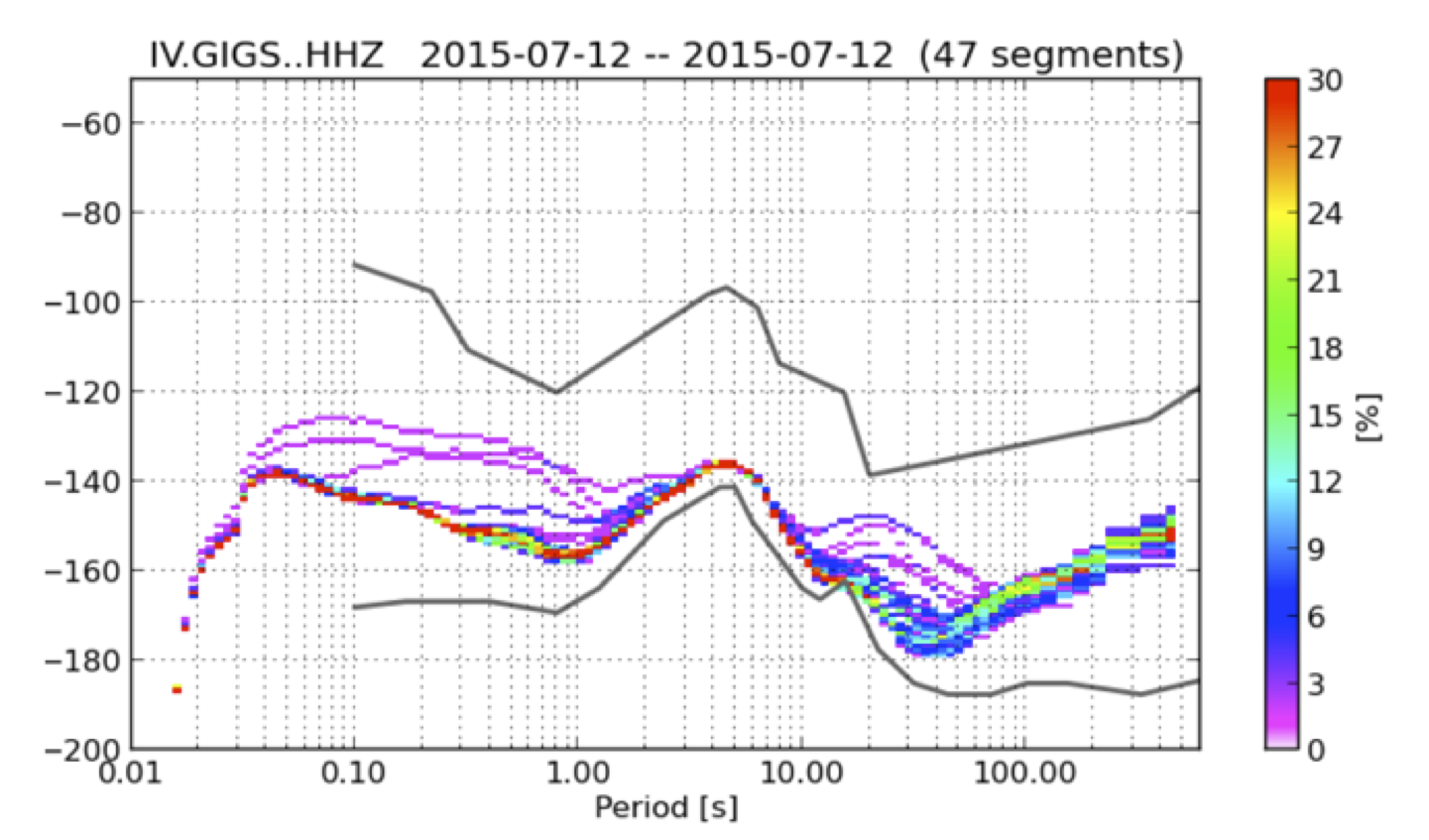} 
 \caption{Typical power spectrum of the vertical component.}
 \label{seism1}
 \end{figure}

\begin{figure}
 \centering
 \includegraphics[width=12 cm]{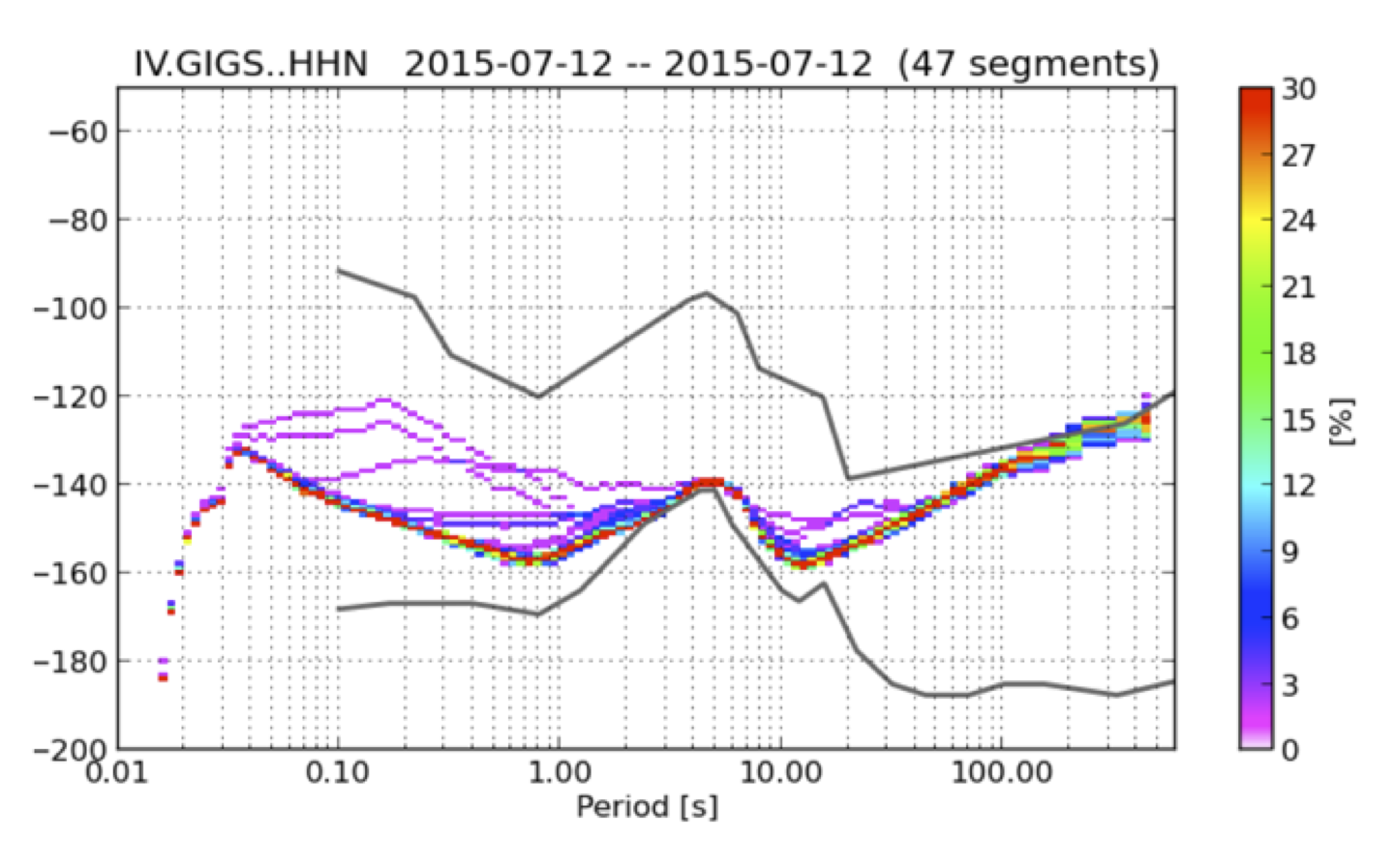} 
 \caption{Typical power spectrum of the horizontal N component.}
 \label{seism2}
 \end{figure}

\begin{figure}
 \centering
 \includegraphics[width=12 cm]{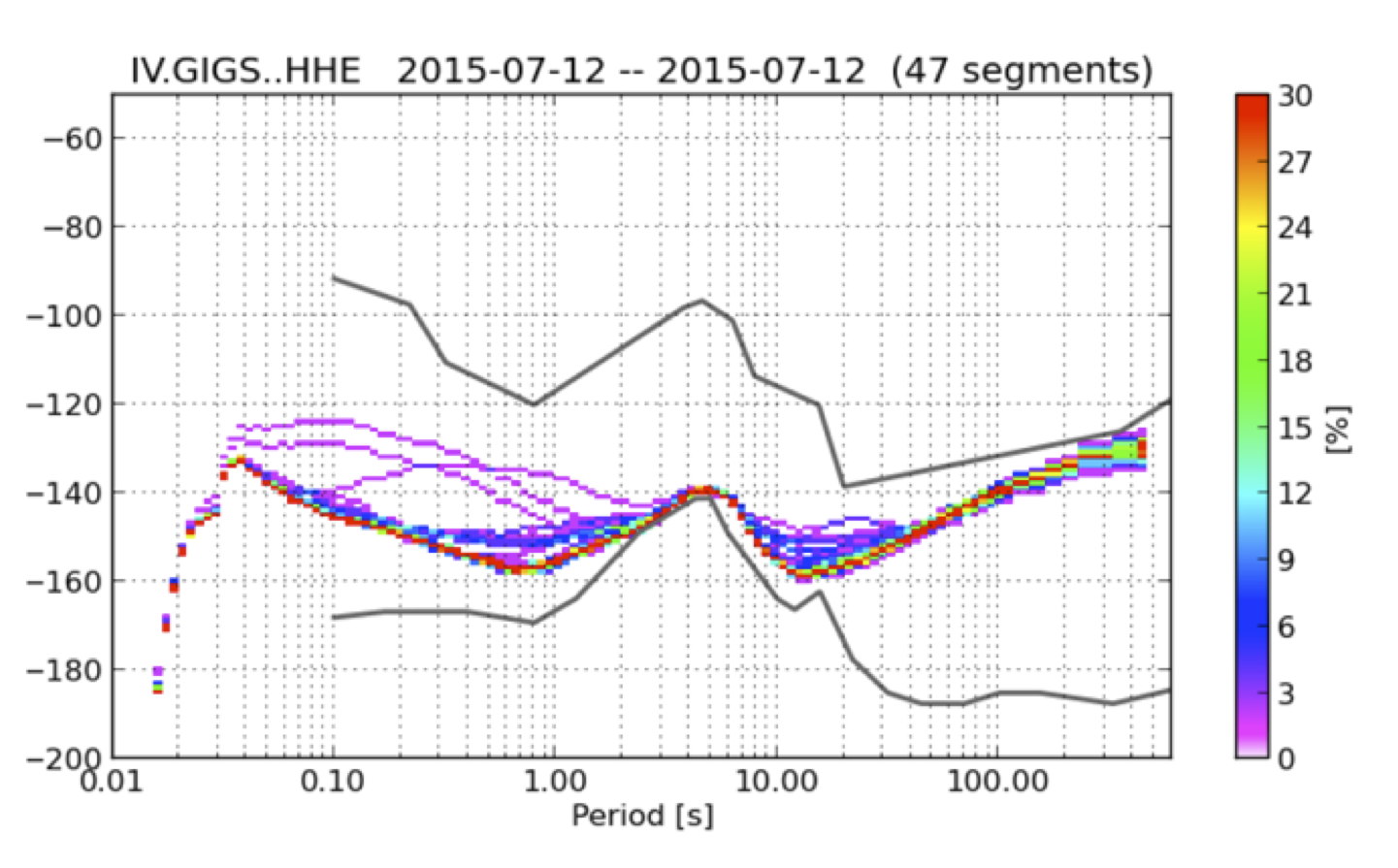} 
 \caption{Typical power spectrum of the  horizontal E component.}
 \label{seism3}
 \end{figure}

\begin{figure}
 \centering
 \includegraphics[width=12 cm]{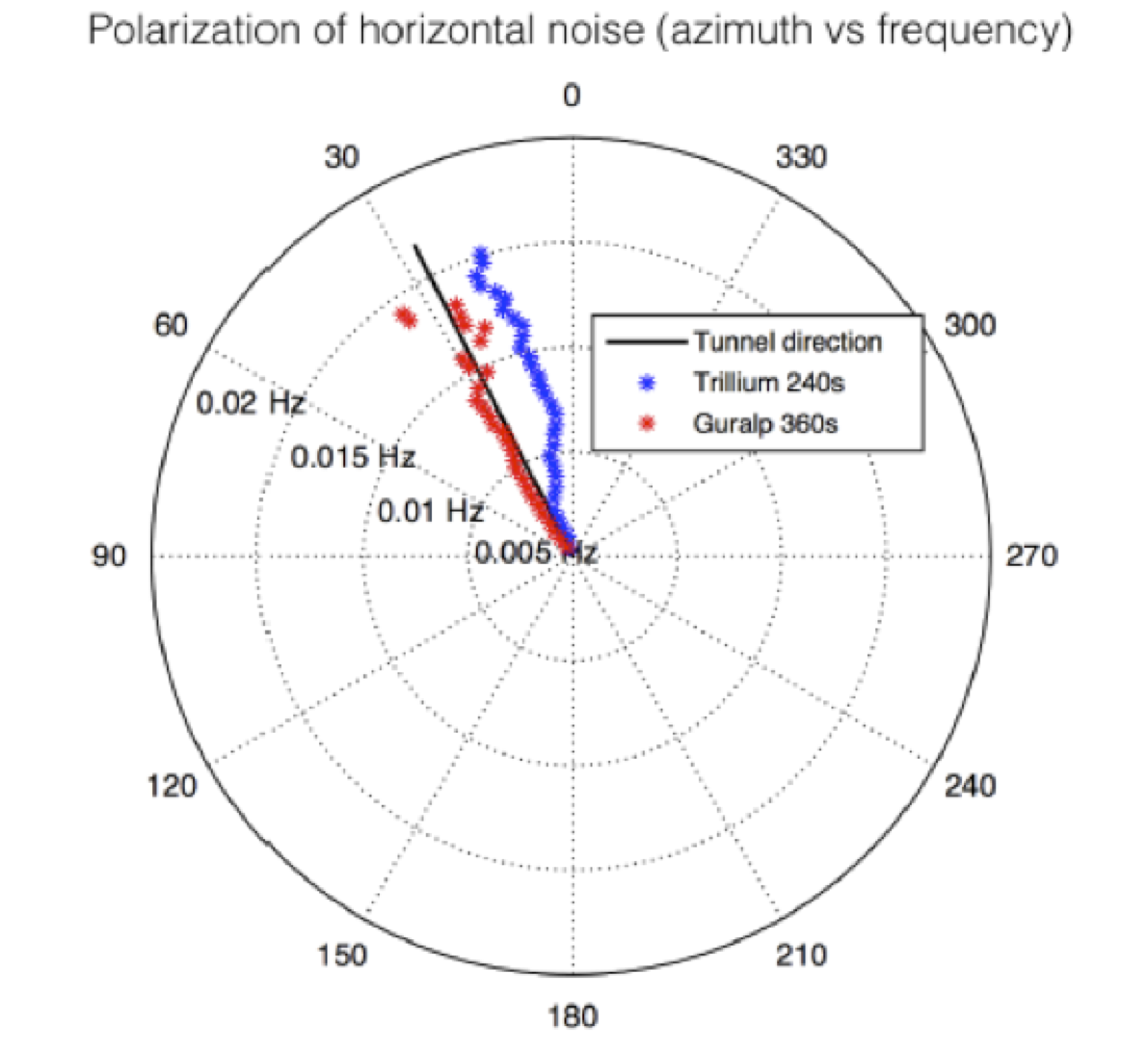} 
 \caption{Reconstruction of the polarization of the horizontal noise of the two seismometers. Their relative alignment is within 5 degrees, they show the disturbances propagates mainly along the tunnel.}
 \label{polar}
 \end{figure}

\begin{figure}
 \centering
 \includegraphics[width=12 cm]{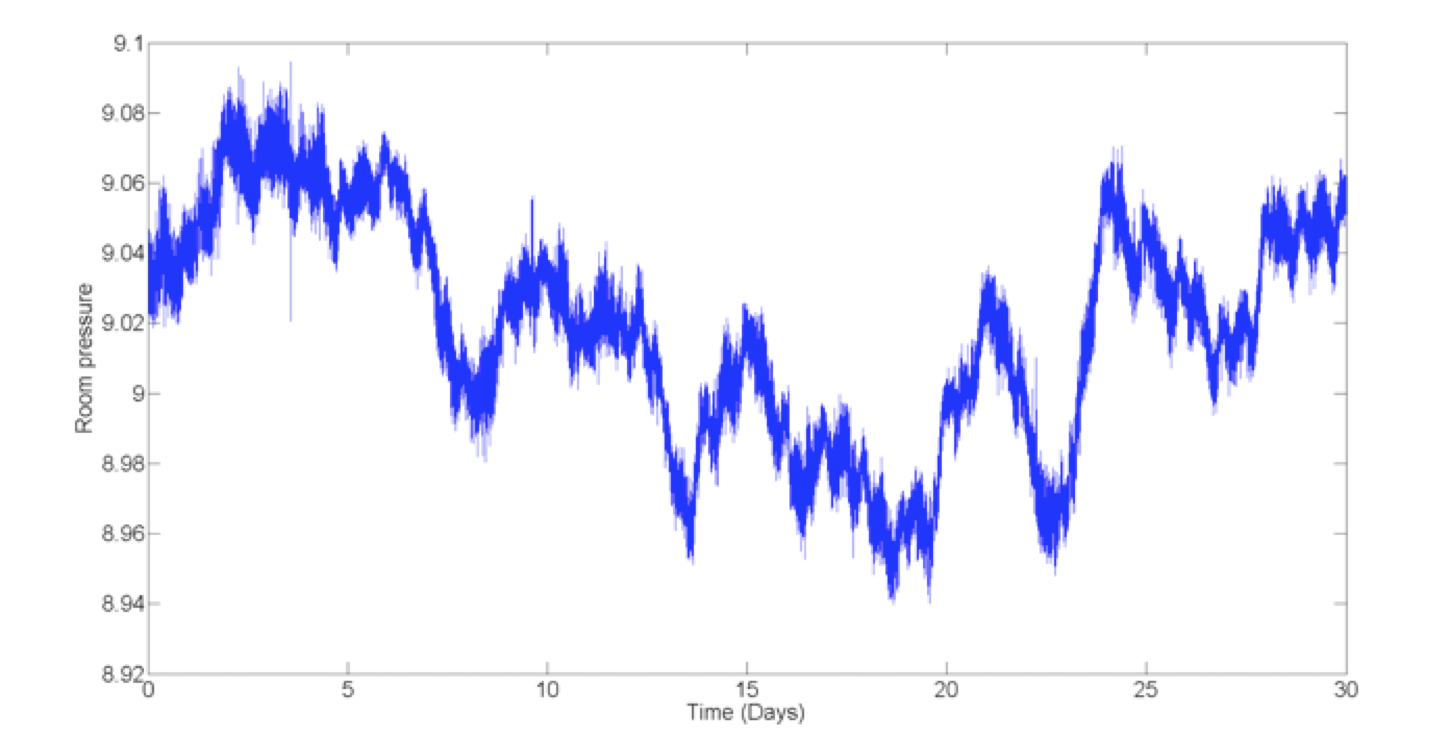} 
 \caption{Typical trend of the pressure inside the room of GINGERino, 30 days. The pressure is express in $kPa/100$.}
 \label{pressure}
 \end{figure}

\begin{figure}
 \centering
 \includegraphics[width=12 cm]{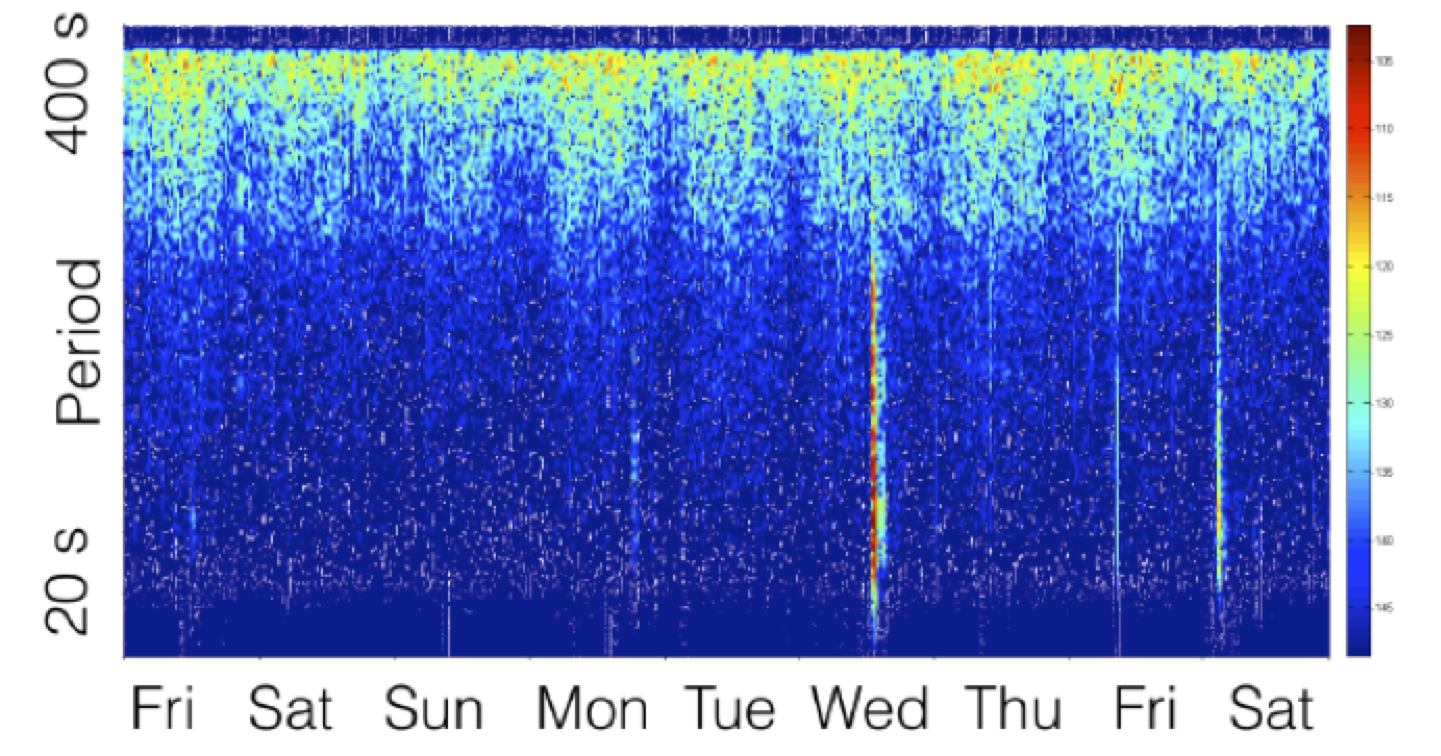} 
 \caption{Seismometers: typical horizontal signals in function of frequency and week day.}
 \label{daynight}
 \end{figure}

In October 2015 we have restarted GINGERino by testing a novel set of mirrors with a measured ring-down time of the cavity around 150ms, about a factor 2 worst than our first run. 
At the same time we successfully implemented the synchronisation with GPS, all the pipes of the ring-laser have been decoupled from the floor, the vacuum pump has been taken away, and the monitor  of the gain tube (required by the backscattering subtraction) has been inserted. In Fig. \ref{comparison}, typical power spectra coming from the two runs are compared.
Thanks to the GPS synchronisation we can perform direct comparison between the RLG data and the seismometers that are independently acquired. 
In particular, we have been able to see the effect of the Vanuatu earthquake happened at UTC time Tuesday, October 20, 2015 21:52 PM (see Fig. \ref{vanuatu}) onto the Sagnac signal.
In order to have clear picture of the residual seismic noise of the site, we have performed a comprehensive analysis of the three seismometers.
The three pictures below (Fig.s \ref{seism1}, \ref{seism2} and \ref{seism3}) show the seismic noise power spectra vs. seismic wave periods, for the three components 
of the acceleration (in units of m${}^{2}$/s${}^{4}$/Hz (dB)): vertical Z, horizontal N (North) and horizontal (East).
The continuous lines are the two spectrum of the low and high noise model for the Earth (NLNM and NHLM, after Peterson 1993 \cite{Peterson}). 
Our typical spectrum is close to the NLNM and shows a very good behaviour across the spectral region for primary and secondary micro-seisms, but exhibits larger and unwanted noise at low frequency (high periods) for the N and E components.
This is a point of concern for the future development of GINGER that is mainly interested at low frequency, and it is necessary to understand if this noise is intrinsic to the lab, or not. 
A deeper analysis including the polarisation for horizontal components of acceleration has shown that the excess horizontal noise is directional and directed along the tunnel (see Fig. \ref{polar}). Following the literature (Beauduin, R., et al. "The effects of the atmospheric pressure changes on seismic signals 
or how to improve the quality of a station." Bulletin of the Seismological Society of America 86.6 (1996): 1760--1769), a possible explanation is that it is induced by the air motion around the seismometers.
 Further investigation will be necessary, but it is important to show that  the environmental pressure monitors has shown a large variability over time (see Fig. \ref{pressure}). If the further analysis will show that the pressure variability is causing the problem, the  solution will be to further isolate the apparatus with pressure tight doors or similar protections.
Day/Night and weekdays/week-end variations are evident and very well correlated with the TF analysis of the seismic noise on horizontal component parallel to the tunnel. In Fig. \ref{daynight} the analysis done for days of year from 164 to 172 is reported.
\section{Conclusions}
GINGERino has been constructed inside LNGS and at the moment it is in the commissioning phase. Several mechanical improvements have been done in order to make the RLG cavity stiffer 
and to improve its isolation from external disturbances.
The instrument has been in data taking during the spring and in October 2015, so far. In the first run the ring-down time of the cavity was of the order of $240 \mu s$, 
the second one was about $150\mu s$, the quality of the cavity depends on the mirrors. A novel set of higher quality mirror is expected to be in place before summer 2016. 
The sensitivity curve in both cases shows a level around $10^{-10}$ $rad/s$ compatible with the actual instrument shot noise and ringdown times.
In 2016 the effort will be dedicated to increase the ringdown time of the cavity, possibly up to $1ms$, in this way the sensitivity will increase by a factor $2$ and the backscattering will be as well reduced. 
Notwithstanding this technical limitation GINGERino has already given few clear indication on the site quality toward the bigger GINGER project.
The analysis of RLG data and co-located seismometers and other environmental monitors has shown that pressure is the parameter with higher variability. 
A first evidence of the role of this variations on the low frequency disturbances seen by the seismometer has been found.
A possible solution to this problem is to isolate the ring in a room equipped with hermetic doors in order to maximally reject noise coming from air currents, fans etc. 
We will investigate in the near future if this solution can be pursued in LNGS. 
Few tele seismic events have been detected by GINGERino, coming from far away earthquakes; the analysis of those events are in progress.
\section*{Acknowledgement}
The construction of GINGERino has been a very tough job, which has been possible for the big effort of several colleagues of Pisa, Naples and LNGS. For that we are grateful to:
A. Soldani, G. Petragnani, G. Balestri, M. Garzella, A. Sardelli and F. Francesconi of Pisa, and   G. Bucciarelli, Lorenzo Marrelli and Nicola Massimiani (Servizio Esercizio e Manutenzione) and Francesco Esposito, Luca Faccia, Giustino D'Alfonso, Stefano Giusti e Luigi Rossi of 'Servizio Facchinaggio' at LNGS. A special thanks goes to C. Zarra for her continuous assistance at LNGS and to Pino Passeggio of Naples, who has taken care of the acoustic box. We have to thank as well the Computing and Network teams of LNGS and Pisa Section for its continuous assistance. We have to acknowledge the very professional support given by Laurent Pinard of the Laboratoire des Mat\'eriaux Avanc\'es (CNRS-Lyon), who has  given us the necessary support for all concerning the mirrors.
\section*{References}
\bibliography{mybib}{}

\begin{thebibliography}{10}
\ifx \showCODEN  \undefined \def \showCODEN #1{CODEN #1}  \fi
\ifx \showISBN   \undefined \def \showISBN  #1{ISBN #1}   \fi
\ifx \showISSN   \undefined \def \showISSN  #1{ISSN #1}   \fi
\ifx \showLCCN   \undefined \def \showLCCN  #1{LCCN #1}   \fi
\ifx \showPRICE  \undefined \def \showPRICE #1{#1}        \fi
\ifx \showURL    \undefined \def \showURL {URL }          \fi
\ifx \path       \undefined \input path.sty               \fi
\ifx \ifshowURL \undefined
     \newif \ifshowURL
     \showURLtrue
\fi

\bibitem{RSIUlli}
Karl~Ulrich Schreiber and Jon-Paul~R. Wells.
\newblock Invited review article: Large ring lasers for rotation sensing.
\newblock {\em Review of Scientific Instruments}, 84\penalty0 (4), 2013.
\newblock \ifshowURL {\showURL
  \path|http://scitation.aip.org/content/aip/journal/rsi/84/4/10.1063/1.4798216|}\fi.

\bibitem{EarthUlli}
T.~et~al. Nilsson.
\newblock Combining vlbi and ring laser observations for determination of high
  frequency earth rotation variation.
\newblock {\em Journal of Geodynamics}, 62\penalty0 (12):\penalty0 69--73, 12
  2012.
\newblock \showISSN{0264-3707}.

\bibitem{ChandlerUlli}
K.~U. et~al. Schreiber.
\newblock How to detect the chandler and the annual wobble of the earth with a
  large ring laser gyroscope.
\newblock {\em Physical Review Letters}, 107\penalty0 (17), 10 2011.
\newblock \showISSN{0031-9007}.

\bibitem{NoiPRD}
F.~Bosi, G.~Cella, A.~Di~Virgilio, A.~Ortolan, A.~Porzio, S.~Solimeno,
  M.~Cerdonio, J.~P. Zendri, M.~Allegrini, J.~Belfi, N.~Beverini, B.~Bouhadef,
  G.~Carelli, I.~Ferrante, E.~Maccioni, R.~Passaquieti, F.~Stefani, M.~L.
  Ruggiero, A.~Tartaglia, K.~U. Schreiber, A.~Gebauer, and J.~P.~R. Wells.
\newblock Measuring gravitomagnetic effects by a multi-ring-laser gyroscope.
\newblock {\em Physical Review D}, 84\penalty0 (12), 2011.
\newblock \showISSN{1550-7998}.
\newblock \ifshowURL {\showURL \path|<Go to ISI>://WOS:000297938000001|}\fi.

\bibitem{NoiseUlli}
A.~et~al. Gebauer.
\newblock High-frequency noise caused by wind in large ring laser gyroscope
  data.
\newblock {\em Journal of Seismology}, 16\penalty0 (4):\penalty0 777--786, 08
  2013.
\newblock \showISSN{1383-4649}.

\bibitem{simonelli}
A~Simonelli, J~Belfi, N~Beverini, G~Carelli, A~Di~Virgilio, E~Maccioni,
  G~De~Luca, and G~Saccorotti.
\newblock First deep underground observation of rotational signals from an
  earthquake at teleseismic distance using a large ring laser gyroscope,
  January 2016.
\newblock \ifshowURL {\showURL \path|http://arxiv.org/abs/1601.05960|}\fi.

\bibitem{NoiCR}
Angela Di~Virgilio, Maria Allegrini, Alessandro Beghi, Jacopo Belfi, Nicolo
  Beverini, Filippo Bosi, Bachir Bouhadef, Massimo Calamai, Giorgio Carelli,
  Davide Cuccato, Enrico Maccioni, Antonello Ortolan, Giuseppe Passeggio,
  Alberto Porzio, Matteo~Luca Ruggiero, Rosa Santagata, and Angelo Tartaglia.
\newblock A ring lasers array for fundamental physics.
\newblock {\em Comptes Rendus Physique}, 15\penalty0 (10):\penalty0 866--874,
  2014.
\newblock \showISSN{1631-0705}.
\newblock \ifshowURL {\showURL \path|<Go to ISI>://WOS:000348002300006|}\fi.

\bibitem{PhDRosa}


\bibitem{NoiRosa}
R.~Santagata, A.~Beghi, J.~Belfi, N.~Beverini, D.~Cuccato, A.~Di~Virgilio,
  A.~Ortolan, A.~Porzio, and S.~Solimeno.
\newblock Optimization of the geometrical stability in square ring laser
  gyroscopes.
\newblock {\em Classical and Quantum Gravity}, 32\penalty0 (5), 2015.
\newblock \showISSN{0264-9381}.
\newblock \ifshowURL {\showURL \path|<Go to ISI>://WOS:000350637200013|}\fi.

\bibitem{NoiAPB}
J.~Belfi, N.~Beverini, F.~Bosi, G.~Carelli, A.~Di~Virgilio, E.~Maccioni,
  A.~Ortolan, and F.~Stefani.
\newblock A 1.82 m(2) ring laser gyroscope for nano-rotational motion sensing.
\newblock {\em Applied Physics B-Lasers and Optics}, 106\penalty0 (2):\penalty0
  271--281, 2012.
\newblock \showISSN{0946-2171}.
\newblock \ifshowURL {\showURL \path|<Go to ISI>://WOS:000299749700005|}\fi.

\bibitem{NoiVirgo}
Jacopo Belfi, Nicolo Beverini, Filippo Bosi, Giorgio Carelli, Angela
  Di~Virgilio, Dmitri Kolker, Enrico Maccioni, Antonello Ortolan, Roberto
  Passaquieti, and Fabio Stefani.
\newblock Performance of "g-pisa" ring laser gyro at the virgo site.
\newblock {\em Journal of Seismology}, 16\penalty0 (4):\penalty0 757--766,
  2012.
\newblock \showISSN{1383-4649}.
\newblock \ifshowURL {\showURL \path|<Go to ISI>://WOS:000308328000017|}\fi.

\bibitem{NoiMetrologia}
D.~Cuccato, A.~Beghi, J.~Belfi, N.~Beverini, A.~Ortolan, and A.~Di~Virgilio.
\newblock Controlling the non-linear intracavity dynamics of large he-ne laser
  gyroscopes.
\newblock {\em Metrologia}, 51\penalty0 (1):\penalty0 97--107, 2014.
\newblock \ifshowURL {\showURL \path|<Go to ISI>://WOS:000331206100016|}\fi.

\bibitem{Hurst:14}
Robert~B. Hurst, Nishanthan Rabeendran, K.~Ulrich Schreiber, and Jon-Paul~R.
  Wells.
\newblock Correction of backscatter-induced systematic errors in ring laser
  gyroscopes.
\newblock {\em Appl. Opt.}, 53\penalty0 (31):\penalty0 7610--7618, Nov 2014.
\newblock \ifshowURL {\showURL
  \path|http://ao.osa.org/abstract.cfm?URI=ao-53-31-7610|}\fi.

\bibitem{CuccatoT}
D.~Cuccato.
\newblock Modeling, estimation and control of ring laser gyroscopes for the
  accurate estimation of the earth rotation.
\newblock 2015.
\newblock \ifshowURL {\showURL
  \path|http://www.infn.it/thesis/thesis_dettaglio.php?tid=10072|}\fi.

\bibitem{Isogai13}
T.~Isogai, J.~Miller, P.~Kwee, L.~Barsotti, and M.~Evans.
\newblock Loss in long-storage-time optical cavities.
\newblock {\em Opt. Express}, 21\penalty0 (24):\penalty0 30114--30125, Dec
  2013.
\newblock \ifshowURL {\showURL
  \path|http://www.opticsexpress.org/abstract.cfm?URI=oe-21-24-30114|}\fi.

\bibitem{Sridhar2011}
G.~Sridhar, Sandeep~K. Agarwalla, Sunita Singh, and L.~M. Gantayet.
\newblock Cavity ring-down technique for measurement of reflectivity of high
  reflectivity mirrors with high accuracy.
\newblock {\em Pramana}, 75\penalty0 (6):\penalty0 1233--1239, 2011.
\newblock \showISSN{0973-7111}.
\newblock \ifshowURL {\showURL
  \path|http://dx.doi.org/10.1007/s12043-010-0211-8|}\fi.

\bibitem{Peterson}
J.~Peterson.
\newblock Observations and modeling of seismic background noise.
\newblock {\em U.S. Geol. Surv. Open-File Rept.}, 93\penalty0 (322), 1993.
\newblock \showISSN{1-95}.

\end{thebibliography}
\bibliographystyle{is-unsrt}
\end{document}